\documentclass[aps,superscriptaddress,prb,twocolumn,floatfix]{revtex4}
\usepackage{natbib}
\usepackage{graphicx}
\usepackage{dcolumn}
\usepackage{bm}
\usepackage{amssymb,amsmath,txfonts}
\begin{document}
\preprint{APS/123-QED}
\title{Experimental Verification of the Quantized Conductance of Photonic Crystal Waveguides}
\author{W. Dai}
\affiliation{Ames Laboratory-USDOE, and Department of Physics and
     Astronomy, Iowa State University, Ames IA, 50011}
\author{B. Wang}
\affiliation{Ames Laboratory-USDOE, and Department of Physics and
     Astronomy, Iowa State University, Ames IA, 50011}
\author{Th. Koschny}
\author{C. M. Soukoulis}
\affiliation{Ames Laboratory-USDOE, and Department of Physics and
     Astronomy, Iowa State University, Ames IA, 50011}
\affiliation{Institute of Electronic Structure and Lasers (IESL), FORTH, and
   Department of Material Science and Technology, University of Crete, 71110
   Heraklion, Crete, Greece.}

\date{\today}
 
\begin{abstract}
We report experiments that demonstrate the quantization of the conductance
of photonic crystal waveguides. To obtain a diffusive wave, we have added all
the transmitted channels for all the incident angles. The conductance steps
have equal height and a width of one half the wavelength used. Detailed
numerical results agree very well with the novel experimental results.
\end{abstract}

\pacs{Valid PACS appear here}
\maketitle

Quantization of different physical quantities is one of the
interesting phenomena in science. The quantization occurs because of
the wave nature of particles. One quantity that gives strong
quantization is the electrical conductance $g$, the inverse of the
resistance. It is very difficult to calculate the transport properties
of small devices analytically. Landauer was the first one to make the
connection between the conductance and the transmission coefficient
\cite{Lan}. He showed that $G=(e^2/\hbar\pi)T/(1-T)$, where $T$ is the
transmission coefficient. If one has a perfect metal $T=1$, then
$G\rightarrow\infty$ and therefore $R\rightarrow 0$, as expected for a
perfect metal. In 1981, Soukoulis and Economou \cite{souk} proved, by
using the Kubo-Greenwood formula, that for 1D systems
$G=(e^2/\hbar\pi)T$. There was a fierce controversy in 1980's
regarding which formula was correct \cite{cont}, since the
Economou-Soukoulis formula gives a finite value for the resistance for
the perfect metal. The experiments \cite{e1,e2} resolved this issue
and indeed it was found that $G=(e^2/\hbar\pi)T$. Extensions to higher
dimension \cite{hd1,hd2} were achieved for G and it was shown that for
a multichannel wire the dimensionless conductance $g$

\newcommand{\abs}[1]{\lvert#1\rvert}
\begin{equation}
  g=G\biggl/\biggl(\frac{e^2}{\hbar\pi}\biggr)=\sum_{n,m=1}^{N_c}\abs{t_{nm}}^2
  \label{e1}
\end{equation}
where $t_{nm}$ is the transmission coefficient between
incident mode $n$ and output mode $m$. The same set of modes $N_c$ are
used for the incident and output modes.  For classical waves the total
$T$ is equivalent to the dimensionless conductance \cite{prl-1997} in
electronic systems $T=g$. For the ideal waveguide configuration, $g$
gives the number of propagation modes inside the waveguide.

The key idea of Landauer was to relate the resistance of the sample
with its transmission. So the idea of quantization of conductance can
be also obeyed by all types of waves, electromagnetic waves, acoustic
and elastic waves. It is amazing that the only experiment \cite{exp}
that has been done with waves is the transmission of a slit of
variable width for a given wavelength of $\lambda=1.55 \ \mu
m$. Similar to its electronic counterpart, the optical conductance of
a structure is described as the total light transmitted through the
structure from a diffusive illumination (an isotropic incoherent
incident wave). So the conductance $g$ for classical waves is
dimensionless. In the experiment of Montie \textit{et al.}  \cite{exp}
a two dimensional (2D) diffuser was used to achieve the diffusive
illumination. The diffuser was essentially a 2D random array of
scatterers through which the normally incident plane wave scatters
diffusively and isotropically. The diffused light passed through a
metal slit and the transmitted light was collected. The result showed
that the optical conductance increases in a staircase fashion. A new
step occurs when the slit width $W=n\lambda/2$ ($n=1, 2, 3,...$)
\textit{i.e.}, a new mode is enabled in the slit.

Photonic crystals (PCs) can be designed to have a bandgap that
prohibits wave propagation in a certain frequency range
\cite{PC-book}. Line defects of PCs can confine light with frequencies
within the bandgap inside the channel and act as waveguides. PC
waveguides have been studied extensively and many applications have
been proposed \cite{PC-book}. However,
the concept of optical conductance was not applied to PC waveguides
until recently \cite{apl-2007, physicaB} and no experimental work has
been reported to our knowledge. In this letter we study the optical
conductance of photonic crystal (PC) waveguides, both numerically and
experimentally. We will demonstrate experimentally that the optical
conductance of a 2D PC waveguide in microwaves has the similar
staircase effect as the metal slit.


We design the working frequency of the waveguide at microwave region,
which makes the PC easy to be fabricated. One of the main experimental
difficulties is the need of diffusive waves. It is easy to be achieved
at optical wavelengths by a diffuser. For microwaves, a large enough
random array of scatters are needed to act as a diffuser. Since the
wavelength is now of the order of $cm$, the diffuser must be really
big in size and the intensity of the transmitted waves will be too
weak to be used efficiently. So we apply Eq.~(\ref{e1}) in
experiments for classical waves. We measure the transmitted power for
each incident plane wave with different incident angle separately and
sum them up to obtain the conductance of EM waves \cite{apl-2007}.

\begin{figure}
\begin{center}
 \includegraphics[width=0.45\textwidth]{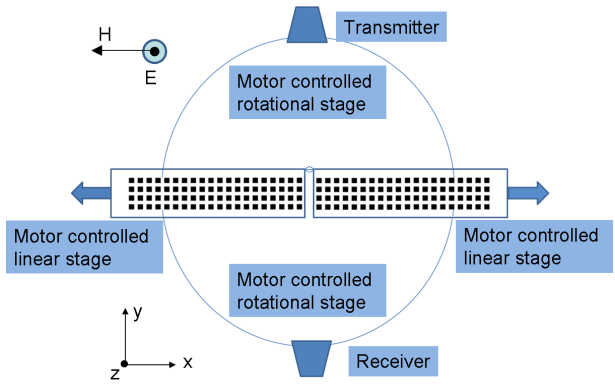}\end{center} 
\caption{An illustration of the experimental setup.}
\label{oc_setup}
\end{figure}

The PC we study is a 2D square array of square alumina rods. The
lattice constant is $a=11$ mm and the square rods are of dimension
$d=3.18$ mm with relative permittivity $\epsilon=9.8$ and height $h=15$ cm. It
has a bandgap between $9.43$ GHz and $12.78$ GHz for TM modes
(Electric field parallel to the rods). The waveguide is formed by two
closely separated pieces of PC slabs with a size of $25a \times 4a$
each. The width of the waveguide ($W$) can be varied by changing the
distance between the two PC slabs. HP8510B network analyzer and a pair
of horn antennas are used to measure the transmission. The antennas
are mounted on motorized rotational stages which can move along a circle
with a radius of $30$ cm and centering at the middle of the entrance of
the waveguide. So both the incident and outgoing angle can be
controlled. (See Fig. \ref{oc_setup} for details).

\begin{figure}
\begin{center}
 \includegraphics[width=0.45\textwidth]{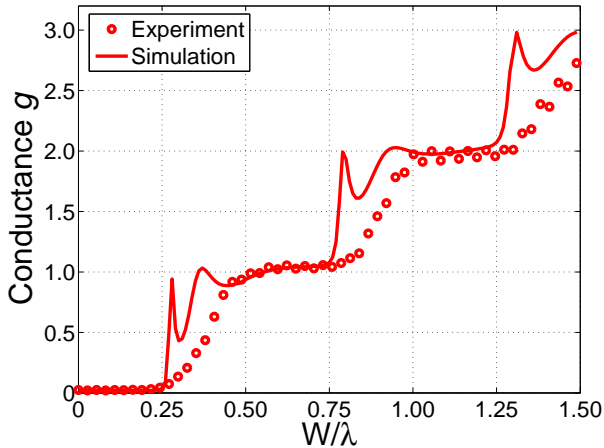}\end{center} 
\caption{(Color online) The experimental and numerical results of
  optical conductance 
of a PC waveguide at $10$ GHz.}
\label{pc_10GHz}
\end{figure}

The experimental result of the optical conductance of the PC waveguide
is shown together with numerical simulations in
Fig. \ref{pc_10GHz}. We see that the optical conductance increases in
a staircase manner with the increase of the waveguide width. The steps
forming the staircases have essentially the same height and a new step
appears at each integer multiple of $\lambda/2$. Notice that the first
conductance step doesn't start at $W/\lambda=0.5$. This is because
that for the photonic crystal channel the boundaries are not well
defined and the field decays exponentially inside the photonic crystal.
So the waveguide is wider than $W$.

\begin{figure}
  \centering
  \includegraphics[width=0.45\textwidth,trim=0 0 0 0,clip]{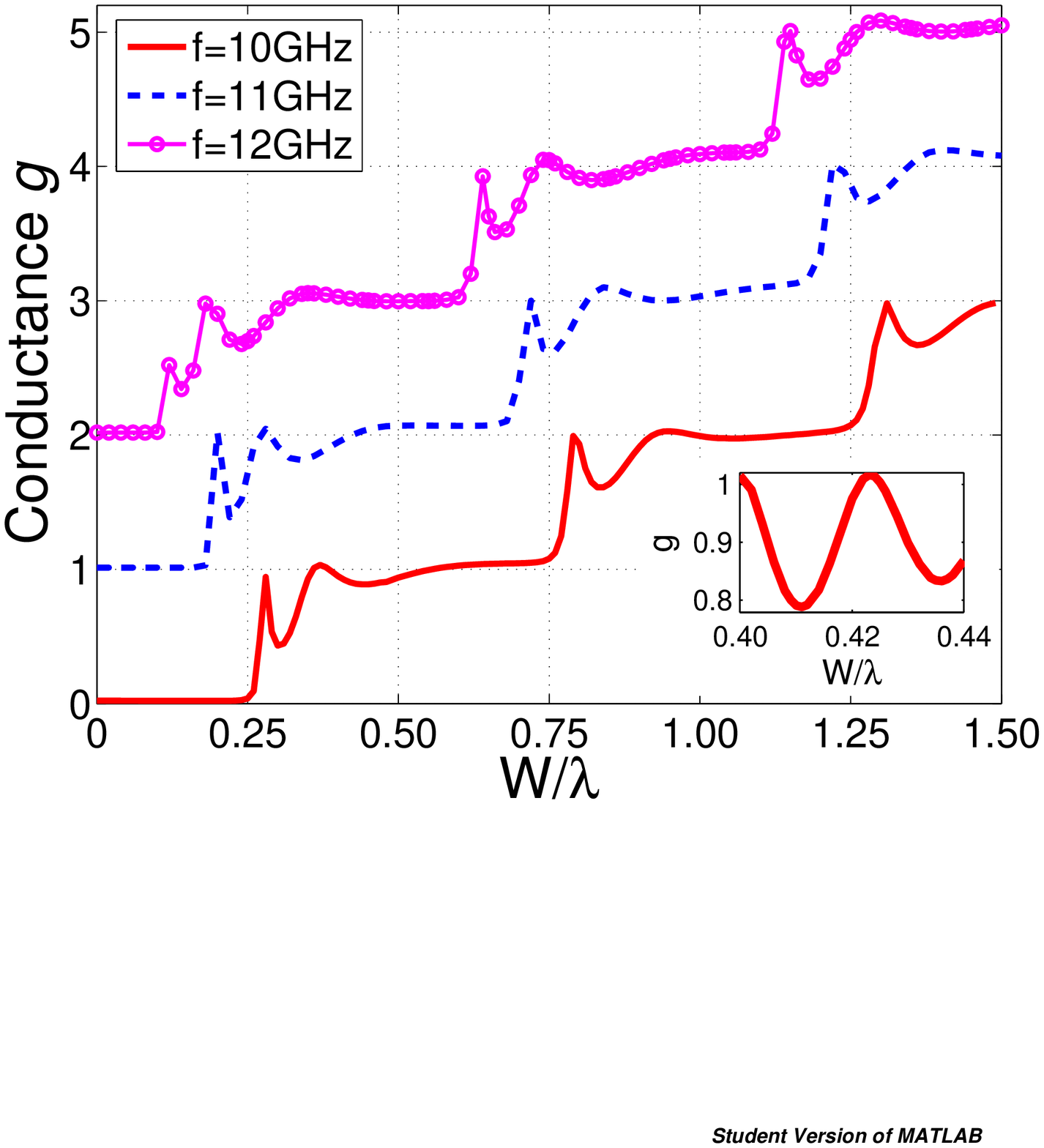}
  \caption{(Color online) The numerical results of optical
  conductance. The working frequencies are 10 GHz, 11 GHz and 12
  GHz. The curves of 11 GHz and 12 GHz are shifted vertically. The PC
  has 4 layers along the channel direction. Inset: $g$ \textit{vs}
  $W/\lambda$ when $f=10$ GHz and the PC has 40 layers.}
  \label{ocpc}
\end{figure}

\begin{figure}
  \centering
  \includegraphics[width=0.45\textwidth]{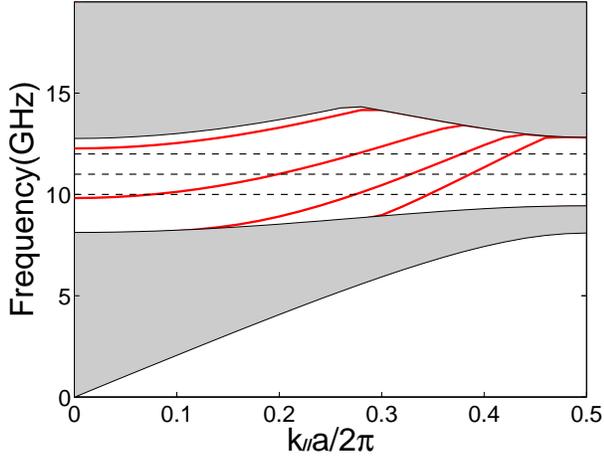}
  \caption{(Color online) Band structure of PC with the channel. The
  channel width is $W=38.673mm$. The dark area is the band area of a
  perfect PC. The dash lines shows three working frequencies. The red
  solid curves shows the impurity bands in the band gap of a perfect
  PC.}
  \label{band}
\end{figure}

\begin{figure}
  \centering
  \includegraphics[width=0.45\textwidth,trim=0 0 0 0,clip]{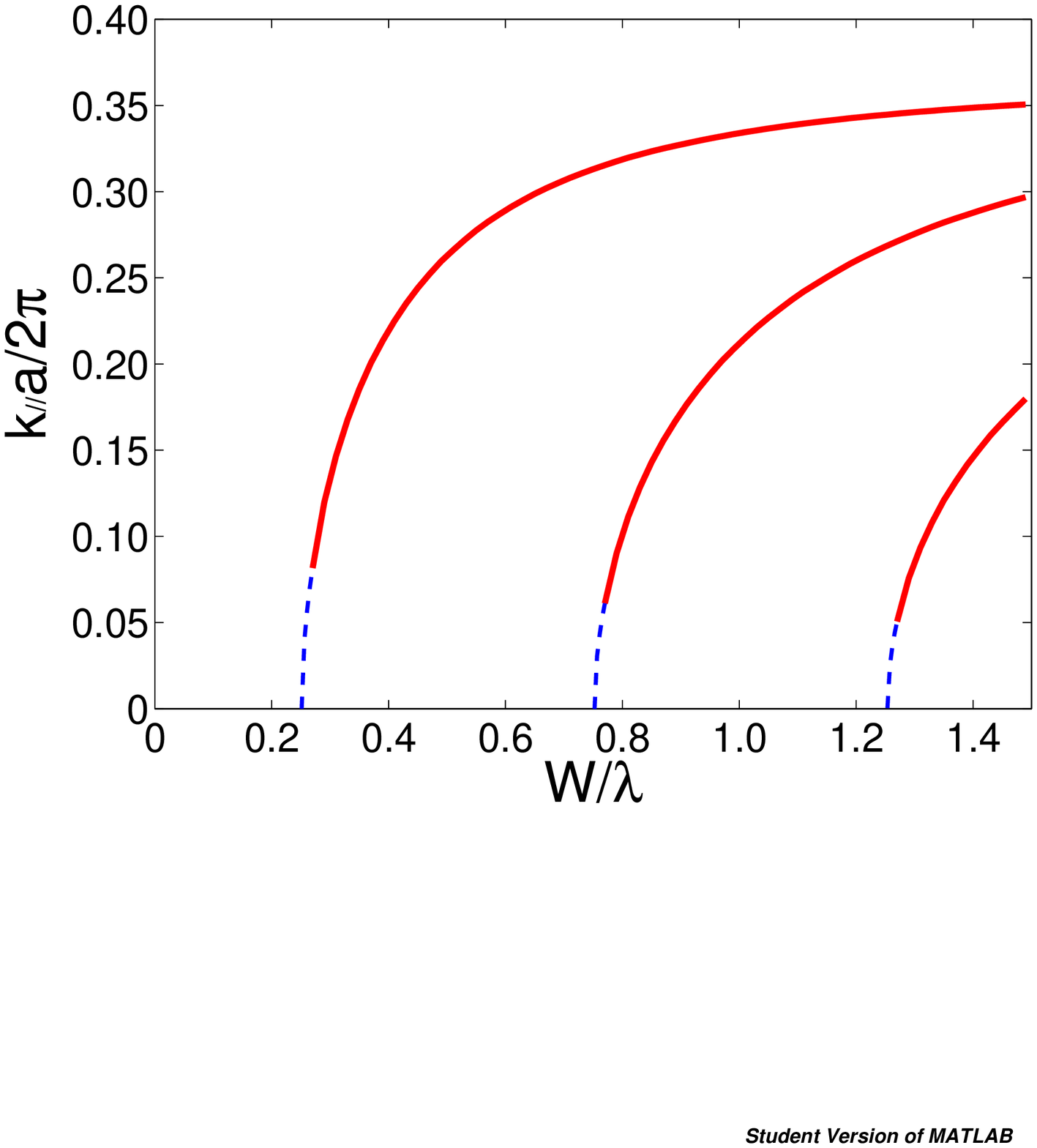}
  \caption{(Color online) Wave vectors of the propagation modes
  \textit{vs.} the width of the channel. The solid lines are gotten
  from band structure of PC with the channel; The dash lines are
  extrapolation of the solid lines. }
  \label{kp}
\end{figure}

We use the commercial software COMSOL Multiphysics to do the numerical
simulations. The simulation results of three frequencies are shown in
Fig. \ref{ocpc}. To understand the shape of optical conductance
curves, the band structure of the PC with the channel are calculated
using the supercell technique. When the channel width increases, some
bands move downwards into the bandgap of the PC. Fig.~\ref{band} shows
the band structure when $W=38.673$ mm. There are four impurity bands
inside the band gap.  The experimental frequency $f=10$ GHz crosses
the three impurity bands.  It means the channel can support three
propagation modes. The propagation wave vectors $k_{\varparallel}$ of
the modes can be read from the horizontal axis. Fig.~\ref{kp} shows
the relations between the propagation wave vectors of the propagation
modes and the width of the channel when $f=10$ GHz based on the band
structure simulations.  The figure demonstrates clearly that the first
propagation mode appears when $W/\lambda=0.25$ and one more
propagation mode when the width increase $\lambda/2$. Comparing with
Fig.~\ref{ocpc}, it is clear that the optical conductance steps
describes the number of propagation modes supported by the channel.

The oscillation of the optical conductance curves can also be
explained by the propagation wave vectors of the channel modes. A
natural explanation about the conductance oscillations is the
Fabry-P\'{e}rot interference, the interference between the multiple
reflection of waves inside the waveguide. The phase difference between
two succeeding reflection equals the propagation phase
$2k_{\varparallel}L$ ($L$ is the length of the channel) plus the phase
change of the reflections at the entrance and the exit of the
channel. The reflection ratio changes with the channel width. To prove
that the conductance oscillations are due to the Fabry-P\'{e}rot
interference, we calculated the optical conductance curves for a long
channel ($L=40a, f=10$ GHz). The inset of Fig.~\ref{ocpc} shows the
simulation result. When the channel width changes from $0.41\lambda$
to $0.43\lambda$, the oscillation has a full period. The reflection
ratio is a continuous function of the channel width. So the reflection
ratio will not change too much during the period since the width
doesn't change too much. Then we neglect them and compare the
propagation phases only. When $W/\lambda=0.411,0.423,0.436$, the
conductance curve has a dip, a peak and another dip. The propagation
phases are $2k_{\varparallel}L=36.10\pi,37.09\pi, 38.09\pi$
respectively. The phase differences between a dip and its neighboring
peak are $0.99\pi$ and $1.00\pi$. They agree with the Fabry-P\'{e}rot
explanation.


From Fig. \ref{pc_10GHz} we can see that the onset of the staircases in
experiments and simulations are in good agreement. However, the
Fabry-P\'erot oscillations are missing in the experiments. The major
reason is that, while in simulations the model is perfectly 2D, 
we do not have an ideal 2D system in experiments. The
source we get from a radiating horn antenna is a confined beam. 
When it encounters the PC structure, while the main part is
propagating in xy plane, the beam also scatters out of the
plane. Since the receiving antenna only collects the transmitted power
in the xy plane where it sits, not all the transmitted power is
obtained. At the onset of the staircase, more power is lost due to the
multiple reflection. So the experimental curve is lower than the one
obtained by simulations and the Fabry-P\'erot resonances are missing.

We also studied the optical conductance of a perfect electric
conductor (PEC) channel, which serves as a simplified model of the
photonic crystal waveguides. Analytical calculations starting from
Maxwell's equations \cite{ana1} are performed to understand the
mechanism behind the stairs.

\begin{figure}[t]
\begin{center}
\includegraphics[width=0.45\textwidth,trim=0 0 0 0,clip]{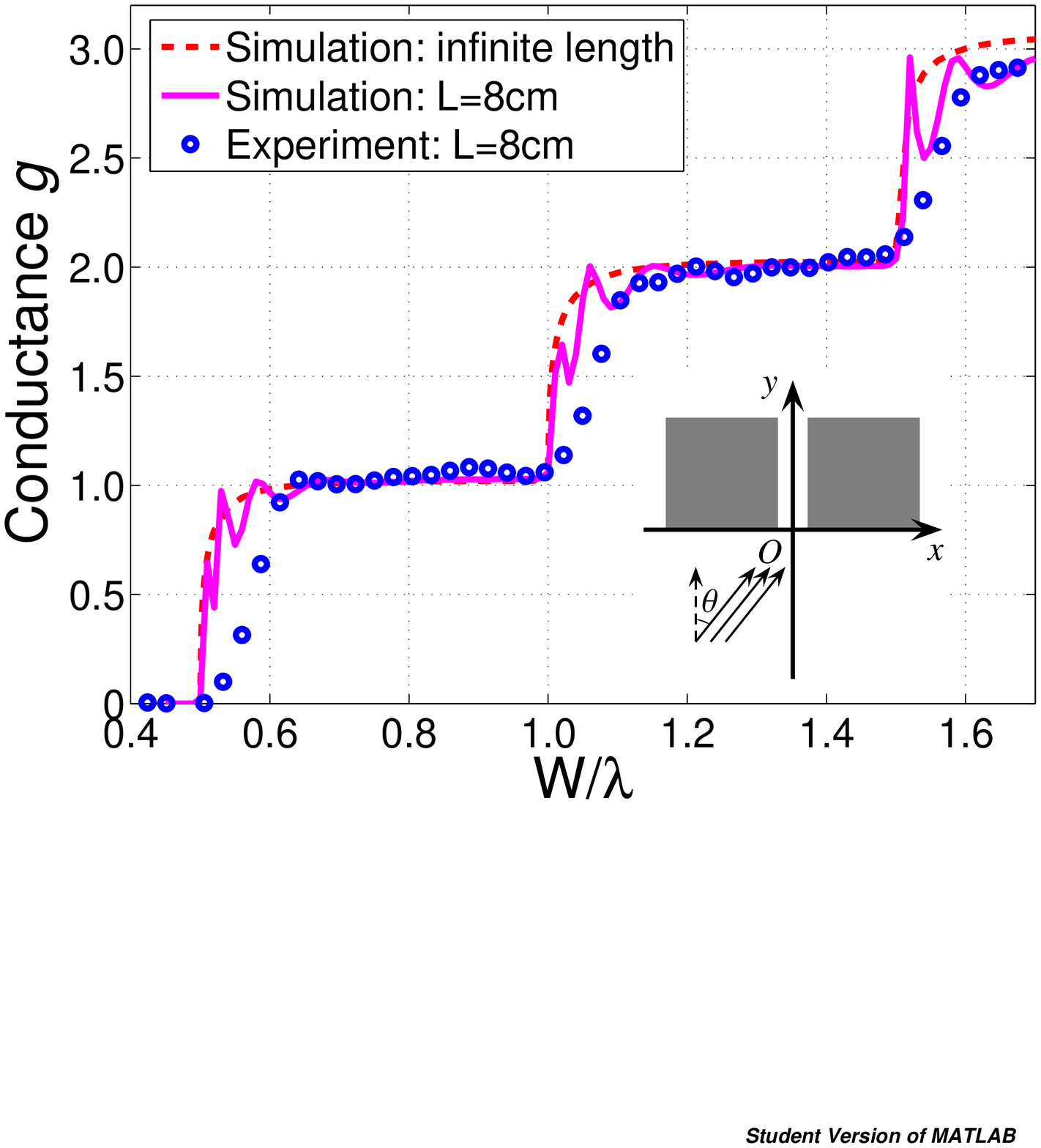}
\end{center} 
\caption{(Color online) The experimental and numerical results of
optical conductance of a PEC channel for TM case. Inset: Diagram of
the PEC channel.}
\label{metal_E}
\end{figure}

Suppose $y=0$ is the interface of the air and the PEC channel. The
half-space $y<0$ is the air and $y>0$ is the PEC channel. $x=0$ is the
middle of the PEC channel (see the inset of Fig.~\ref{metal_E}). 

In TM mode, the incident wave is
\begin{equation}
  E_z^{inc}=\exp[i(k_{\varparallel}y+k_{\bot}x)]. 
\end{equation}
Here $k_{\varparallel}=k_0\cos\theta$; $k_{\bot}=k_0\sin\theta$. $k_0$ is the
wave vector in air; $\theta$ is the incident angle. The reflected wave
is the composition of plane waves. The total electric field in the air
area is
\newcommand{\intii}[0]{\int_{-\infty}^{\infty}}
\begin{equation}
  E_z^{air}=\exp[i(k_{\varparallel}y+k_{\bot}x)]+\intii dk \rho(k) \exp[i(kx-k_yy)].
\label{Ez_air}
\end{equation}
where $k_y=\sqrt{k_0^2-k^2}$, and $\rho(k)$ are unknown coefficients. 

The channel is infinite long; so the field in the channel is the
composition of all the channel modes moving towards $+y$ direction.
\providecommand{\sumi}[1]{\sum_{#1=1}^{\infty}}
\begin{equation}
  E_z^{channel}=\sumi{m} A_m \phi_m(x)\exp[i\beta_m y].
\label{Ez_channel}
\end{equation}
Here $\phi_m(x)=\sin(m\pi(x+W/2)/W)$ is
the channel mode profile and $\beta_m=\sqrt{k_0^2-(m\pi/W)^2}$.

We have defined the fields before the channel (Eq.~(\ref{Ez_air})) and
in the channel (Eq.~(\ref{Ez_channel})) separately. $E_z$ and $H_x$
should be continuous along the interface $y=0$. By using these
boundary conditions we can obtain the following set of equations for
the unknowns $A_n$:

\newcommand{\ky}[0]{\sqrt{k_0^2-k^2}}
\begin{equation}
  A_n\gamma_n\beta_n+\sumi{m}A_mg_{mn}=I_n.
  \label{tmana}
\end{equation}

Here
\newcommand{\intw}[0]{\int_{-W/2}^{W/2}}
\newcommand{\kp}[0]{k_{\bot}}
\newcommand{\kv}[0]{k_{\varparallel}}
\begin{equation}
  I_n=\intw 2\kv \exp[i\kp x]\phi_n(x)dx.
\end{equation}
\begin{equation}
  \gamma_n=\intw \phi_n(x)\phi_n(x)dx=W/2.
\end{equation}
\begin{equation}
\begin{split}
  g_{mn}=&\frac{1}{2\pi}\intii dk \intw dx \intw dx'
      \biggl(
      \ky  \\
      &\phi_m(x')\phi_n(x)\exp[ik(x-x')]
      \biggr).
\end{split}
\end{equation}

This method can be applied to TE mode too.  Because of space
 consrains, we won't repeat it here.

The coefficients of the channel models $A_n$ are solved by
 Equ.~(\ref{tmana}). Then the optical conductance of the infinite long
 channel can be calculated.

\begin{figure}
\begin{center}
\includegraphics[width=0.45\textwidth,trim=0 0 0 0,clip]{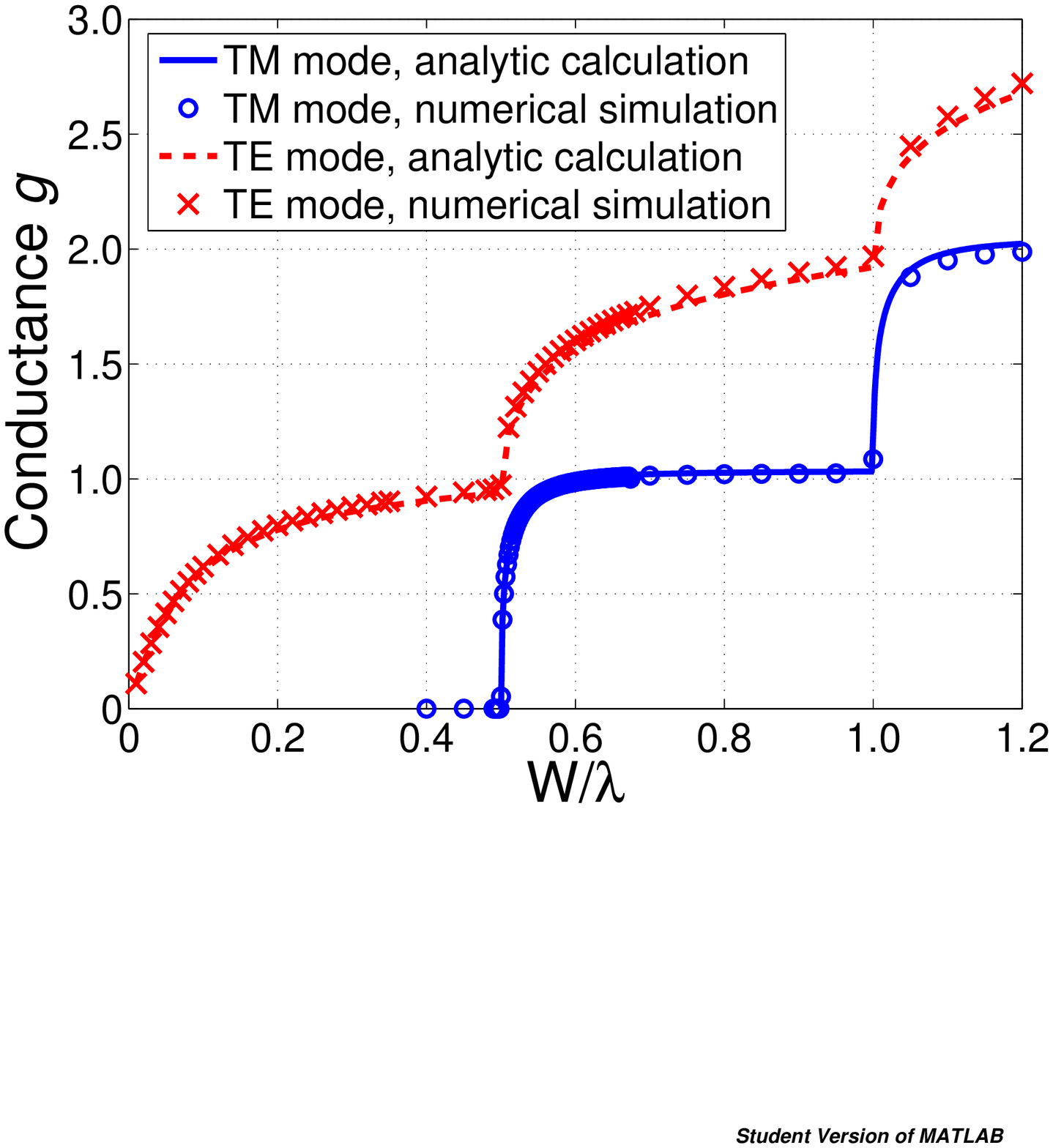}
\end{center} 
\caption{(Color online) Optical conductance of a infinite-long PEC
channel. The first 10 channel modes are chosen in the analytic
calculation; COMSOL Multiphysics is used to do the numerical
simulations.}
\label{ana}
\end{figure}


We also used COMSOL Multiphysics to simulate the infinite long channel
numerically by inserting a perfect matched layer (PML) \cite{pml} in
the channel. The incoming channel modes are absorbed by PML without
reflection. Fig.~\ref{ana} shows the analytic calculation and
numerical simulation results. The Fabry-P\'erot oscillations disappear
since the channel is infinite long. The PEC channel has at least one
TE propagaion mode no matter how narrow channel is. So the first step
of TE mode begins at $W/\lambda=0$. It is an interesting question why
the conductance curves, shown in Fig.~\ref{ana} , of TE and TM modes have
different behavior.

 Experimental measurements have also been performed and the results
are shown together with the simulation results in Fig.~\ref{metal_E}
for TM mode. The experimental curve is smooth and lower than the simulation
results, similar with the PC curves.

We can simplify Equ.~(\ref{tmana}) furthermore. When $k_0W\gg 1$ and
$n\ll k_0W$,
\begin{equation}
  g_{mn}\approx
    \begin{cases}
      \gamma_n\beta_n  \hspace{1cm}&\text{if }m=n;\\
      0                 \hspace{1cm}&\text{if }m\neq n.
    \end{cases}\\
\end{equation}
So $A_n\approx I_n/2\gamma_n\beta_n$.

It means when the channel is wide, we can get the coefficients of the
first several modes directly. Then it is easy to prove that the
contribution to the optical conductance by one mode converges to 1 when
the channel width goes to infinity. This is the reason for the step
profile. 

We have shown for the first time that the conductance of the photonic
crystal waveguide is quantized in microwave frequencies. Quantization
does not only occur in small length scales and in the quantum regime
but also in long length scales and in the classical regime. We have
also introduced new ways to obtain diffusive waves in the microwave
region.

This work was partially supported by Ames Laboratory (Contract
No. DE-AC0207CH11385) and EU projects Metamorphose and Phoremost.


\end{document}